\documentclass{article}
\usepackage{spconf,amsmath,graphicx,booktabs}
\usepackage{amssymb}
\usepackage{caption}
\usepackage{subcaption}

\usepackage{hyperref}
\usepackage[dvipsnames]{xcolor}
\newcommand\myshade{70}
\colorlet{mywholecolor}{MidnightBlue}
\hypersetup{
  linkcolor  = mywholecolor!\myshade!black,
  citecolor  = mywholecolor!\myshade!black,
  urlcolor   = mywholecolor!\myshade!black,
  colorlinks = true,
}

\usepackage{cite}
\newcommand{\minisection}[1]{\vspace{1mm}\noindent\textbf{#1 ---}}

\newcommand{\ve}[1]{\textbf{#1}}
\title{Full-band General Audio Synthesis with Score-based Diffusion}
\name{Santiago Pascual, Gautam Bhattacharya, Chunghsin Yeh, Jordi Pons, Joan Serr\`a}
\address{Dolby Laboratories}
\begin{document}
\ninept
\maketitle
%
%\joan{Use the term ``fullband'' instead of ``full-bandwidth''? See \url{https://en.wikipedia.org/wiki/Wideband_audio} (also applies to the title)}
\begin{abstract}
%Audio synthesis has been primarily developed for speech and music signals. However, 
Recent works have shown the capability of deep generative models to tackle general audio synthesis from a single label, producing a variety of impulsive, tonal, and environmental sounds. Such models operate on band-limited signals and, as a result of an autoregressive approach, they are typically conformed by pre-trained latent encoders and/or several cascaded modules. In this work, we propose a diffusion-based generative model for general audio synthesis, named DAG, which deals with full-band signals end-to-end in the waveform domain. Results show the superiority of DAG over existing label-conditioned generators in terms of both quality and diversity. More specifically, when compared to the state of the art, the band-limited and full-band versions of DAG achieve relative improvements that go up to 40 and 65\%, respectively.
%More specifically, a wideband version of DAG achieves relative improvements upon the state of the art that go up to 40\% in terms of quality and diversity, while the fullband version achieves up to a 65\% improvement. 
%\joan{Also a last sentence of future perspective...}
We believe DAG is flexible enough to accommodate different conditioning schemas while providing good quality synthesis.
\end{abstract}
\section{Introduction}
\label{sec:intro}

Audio synthesis is the computerized generation of audio signals. This task has been primarily tackled under a source-specific paradigm, hence modeling a single type of audio per model. Examples of this paradigm are text-to-speech~\cite{taylor2009text}, music synthesis~\cite{dhariwal2020jukebox,rouard2021crash,nistal2020drumgan}, or less commonly modeled sources like footsteps~\cite{comunita2021footsteps} or laughter~\cite{afsar2021laughter}, among others. 
Nonetheless, a few recent works propose a source-agnostic paradigm, in which generative models can synthesize different types of audio with appropriate conditioning injected into a single neural network. We refer to this as general audio synthesis. 

In this context, Kong et al.~\cite{kong2019samplernn} propose to model environmental sounds with a class-conditioned SampleRNN. This is a deep autoregressive generative model that operates in the  time-domain. % with 16\,kHz sampling rate. \santi{Check whether authors claim 16kHz to be designed due to efficiency} 
On the other hand, Liu et al.~\cite{liu2021pixelsnail} propose another autoregressive model exploiting a VQ-VAE-2 encoder-decoder strategy~\cite{razavi2019vqvae2}. In this setup, a VQ-VAE is trained to create a codebook of audio features extracted from melspectrograms. Hence, the model incorporates a downsampling and upsampling of the original signal space into a discretized latent. Then, a class-conditioned PixelSNAIL autoregressive model~\cite{chen2018pixelsnail} is built as a language model of these discrete latent token sequences. An advantage of this cascaded design is that %in contrast to~\cite{kong2019samplernn}, 
the expensive autoregressive computation can be alleviated while working with a lower time-resolution in the latent, thanks to its auto-encoding design. %Note that this is opposed to the raw signal space modeling of~\cite{razavi2019vqvae2}. 
Since the VQ-VAE operates on melspectrograms, a HifiGAN~\cite{kong2020hifigan} trained with general audio content is plugged on top of the auto-encoder reconstructions to convert them into waveforms. %This modeling strategy operates with 22050\,Hz signals, like HifiGAN's original design~\cite{kong2020hifigan}. 

Other works concurrent to ours tackle more specialized synthesis tasks within the source-agnostic paradigm, like text-to-audio synthesis (that is, generating audio that is representative of a given natural language description). %In this task, a scene is described textually and an audio is generated, which should be representative of the description  ~\cite{yang2022diffsound} \cyeh{this phrase could be removed since people can refer to the citation}. 
This way, following the cascaded framework of Liu et al., Yang et al.~\cite{yang2022diffsound} propose DiffSound, a text-conditioned diffusion probabilistic model that replaces the autoregressive PixelSNAIL and improves generation quality and speed. In addition, Kreuk et al.~\cite{kreuk2022audiogen} propose AudioGen, another cascaded model for text-to-audio in which a discrete latent is learned from raw waveforms and an attention-based autoregressive decoder generates the discrete latent stream. %~\santi{Possibly cite image-to-audio too (Chen et al. 2020), although possibly there is no room for this}.
Note that these cascaded and/or autoregressive designs impose a limitation in the bandwidth of the generated signal. For instance, a component in the cascade can be a band-limited bottleneck. Alternatively, the latent discretization imposed by autoregressive modeling, or the autoregressive design in the raw signal space itself, can lead to sampling rate limitations. This happens to avoid a large quality loss in the compression for the former case, or a high computational inefficiency due to long sequence generations in the latter case.

Despite various advancements on general audio synthesis, we argue that state-of-the-art methods are limited due to (i) targeting audio content below 11\,kHz bandwidth, (ii) reusing previous (and sometimes pre-trained for a different task) modules in a complex cascaded framework, and (iii) quantizing latents with potential quality loss.
%Both DiffSound and AudioGen are also limited in bandwidth, since they operate at 22050\,Hz and 16\,kHz respectively.
In this work, we propose the diffusion audio generator (DAG), a full-band end-to-end source-agnostic waveform synthesizer. While previous works are band-limited due to modeling constraints, DAG is built upon a lossless auto-encoder that can directly generate 48\,kHz waveforms (24\,kHz bandwidth). In addition, its end-to-end design makes it simpler to train and use, avoiding intermediate information bottlenecks or possible cumulative errors from one module to the next one. Furthermore, DAG is built upon a score-based diffusion generative paradigm~\cite{song_generative_2019,song_score-based_2021}, which has shown great performance in related fields like speech synthesis~\cite{kong2020diffwave,chen2020wavegrad}, universal speech enhancement~\cite{serra2022universe}, or source-specific audio synthesis~\cite{rouard2021crash}. Our results show that DAG is capable of generating higher quality content while being a simpler and more parameter-efficient approach. This is evidenced from relative improvements up to 40 and 65\% with respect to the state of the art, depending on whether band-limited or full-band signals are generated. Besides the empirical results reported here, we also provide additional audio samples to showcase some of the possibilities that our solution offers and to highlight its potential.%: \url{https://diffusionaudiosynthesis.github.io}.

% Santi: uncomment this if there is more space
%In the following, we describe our proposed model~(Sec.~\ref{sec:model_proposal}), followed by the experimental setup~(Sec.~\ref{sec:experimental_setup}), where we detail the datasets, baselines and metrics we use. Finally, we discuss results~(Sec.~\ref{sec:results})  and \cyeh{draw} conclusions~(Sec.~\ref{sec:conclusions}). \joan{Consider removing this paragraph if needed (it never adds value)}

\section{Diffusion Audio Generator}
\label{sec:model_proposal}

We propose a deep generative audio model based on score-matching with variance exploding diffusion~\cite{song_generative_2019,song_score-based_2021}. The generator, which follows the denoising score matching strategy from the speech synthesis work~\cite{serra2022universe}, is trained to minimize
\begin{equation*}
    \mathcal{L}_{\text{SCORE}} =  \mathbb{E}_{t} \mathbb{E}_{\ve{z}_t} \mathbb{E}_{\ve{x}_0} \left[ \frac{1}{2} \left\| \sigma_t \tilde{S}(\ve{x}_0+\sigma_t\ve{z}_t,\ve{c},\sigma_t) + \ve{z}_t \right\|_2^2 \right] ,
\end{equation*}
where $t\sim\mathcal{U}(0,1)$, $\ve{z}_t\sim\mathcal{N}(\ve{0},\ve{I})$, $\ve{x}_0\sim p_{\text{data}}$, $\tilde{S}$ is the generated score, $\ve{c}$ is the conditioning signal, and $\sigma_t$ values follow a geometric noise schedule~\cite{song_generative_2019}. We use $\sigma_0=10^{-3}$ and $\sigma_1=1$ in all experiments, which we find sufficient after informal listening. To obtain $\ve{c}$, we project the discrete class labels through an embedding matrix $\mathbf{E} \in \mathbb{R}^{v \times 10v}$, where $v$ is the vocabulary size of the label set. 

To sample audio, we follow noise-consistent Langevin dynamics~\cite{jolicoeur-martineau_adversarial_2021}, corresponding to the recursion
\begin{equation}
\ve{x}_{t_{n-1}} = \ve{x}_{t_n} + \eta \sigma^2_{t_n} \tilde{S}(\ve{x}_{t_n},\ve{c},\sigma_{t_n}) + \beta\sigma_{t_{n-1}}\ve{z}_{t_{n-1}} 
\label{eq:sampling}
\end{equation}
over $N$ uniformly discretized time steps $t_n\in[0,1]$, starting with $\ve{x}_1=\sigma_1\ve{z}_1$. As in~\cite{serra2021tuning,serra2022universe}, we set $\eta$ and $\beta$ with the help of an hyper-parameter $\alpha\in[1,\infty)$: 
\begin{equation*}
\eta=1-\delta^\alpha ,
\quad\quad\quad
\beta = \sqrt{1-\left(\frac{1-\eta}{\delta}\right)^2} ,
\end{equation*}
where $\delta=\sigma_{t_n}/\sigma_{t_{n+1}}$ is the ratio of the geometric progression of the noise. Differently from~\cite{serra2021tuning,serra2022universe}, we here experiment with classifier-free guidance~\cite{ho2022guidance}, which trades-off quality for diversity and introduces another hyper-parameter $\gamma\geq 0$ in the calculation of the score:
\begin{equation*}
    \tilde{S}(\ve{x}_t,\ve{c},\sigma_t) = (1+\gamma)S(\ve{x}_t,\ve{c},\sigma_t) - \gamma S(\ve{x}_t,\sigma_t) ,
\end{equation*}
where $S$ is the output of the DAG network trained with 10\% dropout for unconditional training~\cite{ho2022guidance}. This dropout happens by replacing an actual input label from the training vocabulary by a learnable null token, initialized with zeros, which states that the sample is unconditioned. Since we observed that $\gamma>0$ could yield clipped or distorted results, we decided to rescale the empirically denoised sample~\cite{jolicoeur-martineau_adversarial_2021} (full audio) between $-$1 and 1 at every iteration in case it falls outside that range. Note that this corresponds to a 100$^{\text{th}}$ percentile dynamic thresholding strategy~\cite{imagen_paper}.

\begin{figure}[t]
	\centering
	\includegraphics[width=\linewidth]{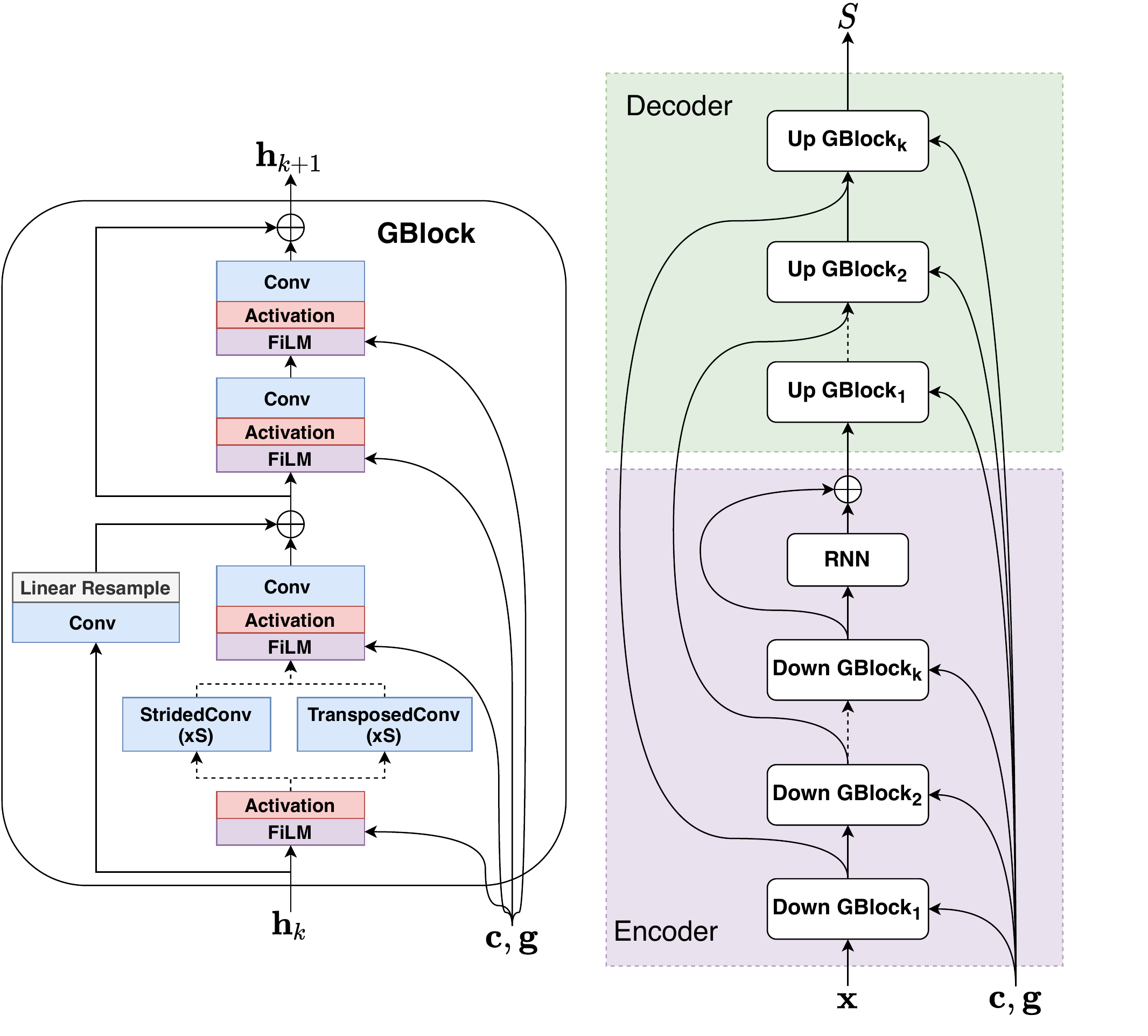}
	\caption{Generator Schematic. Down GBlocks: downsampling encoder blocks (StridedConv). Up Gblocks: decoder upsampling blocks (TransposedConv). The input $\textbf{x}$ is general to the network. Signals $\textbf{h}$ and ($\textbf{c}$, $\textbf{g}$) correspond, respectively, to the input to each GBlock and its conditioning signal's label and $\sigma$ projections.}
	\label{fig:dag}
	%\vspace{2.0cm}
\end{figure}

The DAG architecture is depicted in Figure~\ref{fig:dag}. Inspired by the UNet auto-encoding frameworks of previous score-based diffusion works~\cite{chen2020wavegrad,rouard2021crash,serra2022universe}, the model features an encoder and a decoder. The encoder is built with downsampling convolutional feature encoders. These transform the input signal into a low-rate sequence of hidden feature embeddings through the application of subsequent convolutions with stride factors$~\Delta_k$, where $k$ is the layer index. These factors are configured according the target sampling rate.
Finally, these hidden features are processed through bidirectional gated recurrent units (GRUs), %~\cite{chung2014gru}, 
which aggregate temporal information to attain a much broader receptive field at the decoder input. %This was shown to be beneficial for related audio generative tasks~\cite{serra2022universe}. 
A residual connection surrounding GRUs alleviates possible gradient flow issues due to their saturating activations, and we include a linear projection before the residual summation to adjust dimensionalities.

The decoder converts the encoder hidden features back to the input signal’s resolution by upsampling with transposed convolutions, employing reversed factors with respect to the encoder. In this case, another stack of GBlocks is inserted. The only difference between these blocks and the ones from the encoder are the change of downsampling stages (linear downsampler in skip connection and strided convolution) by upsampling ones (linear upsampler in skip connection and transposed convolution). Also, each encoder level is connected to its corresponding decoder level through a skip connection. Note that all GBlocks feature 4~convolutional blocks, 4~non-linear activations (LeakyReLUs), and 4~FiLM conditioning layers. To obtain the FiLM conditioning signals, we first process the logarithm of $\sigma$ with random Fourier feature embeddings~\cite{rouard2021crash} followed by a multi-layer perceptron~(MLP), which yields the embeddings $\ve{g}$ in Figure~\ref{fig:dag}. Then, $\ve{g}$ is concatenated with $\ve{c}$ to drive the FiLM layers. %~\cite{perez2018film}. 

%Note that DAG's design and the score-based diffusion paradigm allow to infer all time-step predictions in parallel (despite the recursive nature of diffusion and GRUs, the latter operating in the low-rate latent). Therefore, all output time-steps of arbitrary length $T$ after decoding are predicted at once. %This is computationally advantageous with respect to related autoregressive works~\cite{yang2022diffsound}.

\section{Experimental Setup}
\label{sec:experimental_setup}

\subsection{Datasets}
\label{sec:datasets}

To evaluate the effectiveness of DAG to synthesize multiple types of sounds, we resort to two datasets that contain a mixture of impulsive and sustained sound events: UrbanSound8K~\cite{salamon2014us8k} (US8K) and TUT Acoustic Scenes from DCASE 2016 Task~1~\cite{mesaros2016tut} (TUT). US8K contains almost 9~hours of audio, sub-divided into 10~classes that mix ambiance and impulsive sounds, like air conditioner or dog barks. For our experiments, we follow Liu et al.~\cite{liu2021pixelsnail} by first partitioning US8K into 8~hours of training data and 50~minutes of test data. Then, we further divide the training split by taking 10\% of the files for validation across all folds. TUT contains 13~hours of audio, sub-divided into 15~classes, all of them being mainly ambiance sound. The dataset comes with a pre-made training split of almost 10~hours, with 3~hours of test data. As in US8K, we further divide the training partition by taking 10\% of the files for validation.

%is first used to study the behavior of DAG under architectural and sampling configuration variations~\cite{salamon2014us8k}, and as a first point of comparison with respect to the baselines.

\subsection{Baselines and Configuration}
\label{sec:baselines}
%\cyeh{Use "3.2 Configurations" so we can include DAG training setup here}
We compare our approach with two state-of-the-art deep generative models for label-based general audio synthesis: the one of Kong et al.~\cite{kong2019samplernn} (SampleRNN) and the one of Liu et al.~\cite{liu2021pixelsnail} (PixelSNAIL). The former operates in the time-domain, hence performing as an end-to-end solution. The latter operates on pre-trained VQ-VAE latent features. As introduced earlier, both are auto-regressive models and, due to their design constraints, they operate at 16\,kHz and 22050\,Hz, respectively. Here, we re-train both of them at 22050\,Hz.

%\subsubsection{SampleRNN}\label{subsec:samplernn}
\minisection{SampleRNN} Following~\cite{kong2019samplernn}, we implement a two-tier conditional SampleRNN with the previously proposed configuration, using the code from the official repository\footnote{\mbox{\url{https://github.com/qiuqiangkong/sampleRNN_acoustic_scene_generation}}}.  %Its core is built with 3~GRU layers %~\cite{chung2014gru} 
%of 1024~units each, and the MLP on top of it contains two hidden layers of 256~units. %\cyeh{simply cite the authors' implementation is sufficient}. 
%The model is trained as originally on TUT, with our data partitions, 16\,kHz sampling rate, and comparable hyper-parameters. Moreover, we also train a 22050\,Hz version of the model to establish a point of comparison with respect to higher-rate models, which we make confluence at this sampling rate. \joan{If we do not report the 16kHz version remove and adapt these sentences} 
%Experiments were conducted 
The model is trained on our data partition using similar hyper-parameters as in the original publication until we observe convergence (300\,k and 400\,k iterations for US8K and TUT, respectively).
%There is no hyper-parameter change on the model when changing the target sampling rate. 

%\subsubsection{PixelSnail}\label{subsec:pixelsnail}
\minisection{PixelSNAIL} This baseline is based on the three cascaded modules introduced in section~\ref{sec:intro}, which are trained separately: (i) VQ-VAE, (ii) PixelSNAIL, and (iii) HifiGAN~\cite{liu2021pixelsnail}. %Firstly, a VQ-VAE is trained to learn a discretized codebook of acoustic units whose inputs are melspectrogram frames. Secondly, a PixelSNAIL generative model is trained to predict the discrete token sequences that conform label-conditioned audio structures. Thirdly, a HifiGAN is used to emulate neural vocoding in the general audio domain. %\cyeh{we do not need to repeat this description, simply cite and note param changes}. 
Experiments were conducted using the code from the official repository\footnote{\url{https://github.com/liuxubo717/sound_generation}}. The VQ-VAE and PixelSNAIL modules are trained on our data partition with the default hyper-parameters until we observe convergence (100\,k and 500\,k iterations, respectively), and the HifiGAN module is pre-trained on US8K from the repository. For the results on TUT, we also tried to train the HifiGAN module, but observed worse performance (see section~\ref{subsec:comparison}).

\minisection{DAG} DAG is built to operate on full-band audio. In this design, which we name DAG48, we choose $\{2, 2, 4, 4, 5\}$ as $\Delta_k$ stride factors, with block sizes $\{64,128,128,256,512\}$. This way, for 48\,kHz waveforms, we obtain latent sequences of 512~dimensions at 150\,Hz. Nonetheless, we also make DAG operate on 22050\,Hz waveforms (band-limited version) in order to establish a point of comparison with the baselines whilst discarding trivial differences due to bandwidth constraints. We carry out this comparison in two ways. On the one hand, we simply downsample DAG48 generations to 22050\,Hz. On the other hand, we adapt the stride factors to make DAG work directly upon 22050\,Hz signals, with factors $\{2, 2, 3, 3, 5\}$, which approximately correspond to the stride of 7\,ms used in full-band. We name this variant DAG22. Both models are trained for 2\,M and 1.3\,M iterations on US8K and TUT respectively. DAGs are trained via gradient descent with Adam, 8 sequences per batch, and a constant learning rate of $10^{-4}$.

\subsection{Objective Metrics}
\label{sec:obj_metrics}

Previous works evaluate their proposals objectively in terms of quality and diversity~\cite{kong2019samplernn,liu2021pixelsnail}. Their measures rely on auxiliary classifiers to determine how alike does generated data look with respect to real data, or acoustic features where statistics between sets are compared. %These methods require a classifier to be constructed per dataset. 
More recently, Fr{\'e}chet audio distance (FAD) was proposed as a metric to assess source separation quality and was shown to better correlate with human judgement when compared to existing metrics.~\cite{kilgour2019fad}. %This metric is inspired by r{\'e}chet Inception distance~(FID), originally designed to automate the perceptual evaluation of image generation systems~\cite{heusel2017fid}. 
FAD is built upon a VGGish classifier encoder pre-trained on a source-agnostic dataset. Then, the Fr{\'e}chet distance between a reference set and an evaluated set is computed on top of the encoder embeddings. Since this metric is proven to be effective in detecting perceptual nuances in the audio~\cite{kilgour2019fad}, it has been recently used to assess audio generation quality~\cite{rouard2021crash, kreuk2022audiogen}. Nevertheless, a potential limitation of using VGGish is in its input features: the model deals with 16\,kHz signals, which may suffice for classification but is totally insufficient for full-band synthesis.

In this work, we propose to switch from VGGish to an openly available pre-trained audio encoder that operates upon 48\,kHz signals. The motivation is to evaluate the full-band capability of current and future systems performing general audio synthesis. OpenL3 encoders~\cite{cramer2019openl3} are open source front-ends that operate at 48\,kHz (we use the \texttt{env-mel256} front-end with 512-dimensional embeddings from the official repository\footnote{\url{https://github.com/marl/openl3}}). We simply name this metric Fr\'echet distance~(FD), to differentiate it from the previously existing VGGish band-limited approach. Since Fr{\'e}chet distance variants are known to evaluate both quality and diversity in the generated samples
~\cite{naeem2020reliable}, we also consider a logit score~(LS), which follows the Inception score formulation~\cite{barratt2018inception} and focuses the analysis on quality itself. This way, a simultaneously low FD and high LS shows a well functioning model in terms of both quality and diversity. LS is computed upon the logit activations of a pre-trained classifier. Departing from the FD feature extractor, we train an MLP per dataset while freezing the OpenL3 front-end. These MLPs are trained until convergence, and the best scoring iterations in validation are selected. They score 83.1 and 97.2\% on US8K and TUT test sets, respectively.

\section{Results}
\label{sec:results}

\subsection{Comparison with Existing Approaches}
\label{subsec:comparison}

Table~\ref{tab:mainresults} shows the FD and LS scores obtained by all models on US8K and TUT. On the one hand, we observe that PixelSNAIL scores better than sampleRNN on US8K, as reported by Liu et al.~\cite{liu2021pixelsnail}. On the other hand, we observe that retraining all the modules in the PixelSNAIL cascade for TUT (PixelSNAIL w/ TUT-HifiGAN) leads to an undereperforming model, mainly due to issues with HifiGAN. We hypothesize this may be related to the more environmental nature of TUT, and to the HifiGAN losses not being tuned for this scenario. Nonetheless, since HifiGAN operates as a melspectrogram-to-waveform inverter, we can plug the pre-trained HifiGAN from US8K to do the task in TUT. This makes the model perform competitively again with respect to SampleRNN. 

DAG models systematically outperform the baselines and, since the evaluation front-end is pre-trained upon full-band signals, there is a substantial gap between 22050\,Hz and 48\,kHz models, both in terms of FD and LS. The relative gain of DAG48 upon PixelSNAIL in FD is $58.4\%$ on US8K and $64.9\%$ on TUT. In the band-limited scenario, note that only downsampling DAG48 yields 20.5\% and 41\% relative gains for FD and LS, respectively, compared to PixelSNAIL in US8K, which was specifically trained to model the band-limited characteristics of the signal alone. In TUT, the FD improvement is 26.6\%, and for LS it is 9.4\%. However, the DAG22 variant shows that there is room for improving the model performance in band-limited scenarios when training it specifically at 22050 Hz, since the model can score equal or better in FD than DAG48 downsampled. This gap is especially notorious for the TUT dataset. The relative gains for DAG22 upon the baseline then become 24.1\% for FD and 40.0\% for LS on US8K. For TUT, the relative gains are 39.6\% and 18.9\% over PixelSNAIL.
%\textcolor{orange}{The DAG22 variant shows that there may be room for improving the model performance in band-limited scenarios when training it specifically at 22050\,Hz, since the model can score similarly or better in FD, with respect to the downsampled version of DAG48. Although the US8K experiment shows a slightly worse performance, possible due to training fluctuations, the TUT experiment shows a notorious improvement by training the band-limited DAG22 model in the TUT dataset. The relative gains for DAG22 upon the baseline then become 39.6\% for FD and 19\% for LS on TUT over PixelSNAIL.} %For TUT, the relative gains are 39.6\% and 19\% over PixelSNAIL.
%The relative gains for DAG22 upon the baseline then become 24.2\% for FD and 40.7\% for LS on US8K. For TUT, the relative gains are 39.6\% and 19\% over PixelSNAIL. %A more thorough analysis of quality and diversity improvement by type of source can be interesting for future study.

\begin{table*}[h!]
    \centering
    \setlength{\tabcolsep}{8pt}
    \begin{tabular}{lcc c cc c cc}
        \toprule
        \textbf{Model} & \textbf{Samp.~rate} & \textbf{Num.~param.} & & \multicolumn{2}{c}{\textbf{US8K}} & & \multicolumn{2}{c}{\textbf{TUT}} \\
        \cline{5-6}\cline{8-9}
        &  &  & & \textbf{FD}~~$\downarrow$ & \textbf{LS}~~$\uparrow$ & & \textbf{FD}~~$\downarrow$ & \textbf{LS}~~$\uparrow$ \\
        \midrule
        SampleRNN~\cite{kong2019samplernn}            & 22\,kHz    & 41\,M    & & 237.0 & 2.80  & & 138.8 & 2.07 \\
        PixelSNAIL~\cite{liu2021pixelsnail}           & 22\,kHz    & 119\,M   & & 115.1 & 3.27  & & 107.3 & 4.65 \\
        PixelSNAIL~\cite{liu2021pixelsnail} w/ TUT-HifiGAN & 22\,kHz    & 119\,M   & & n/a & n/a  & & 144.3 & 3.33 \\
        \midrule
        DAG48 + Downsample   & 22\,kHz    & 22\,M    & & 91.4  & 5.54  & & 78.8  & 5.13  \\
        DAG22                & 22\,kHz    & 22\,M    & & 87.4  & 5.45  & & 64.8   & 5.74 \\
        DAG48                & 48\,kHz    & 22\,M    & & 47.9  & 6.21  & & 37.7  & 6.59 \\
        \midrule
        Real data            & 22\,kHz    & -     & & 72.6  & 5.26  & & 79.0  & 5.53 \\
        Real data (test set) & 48\,kHz    & -     & & 30.9  & 6.33  & & 3.7  & 9.31 \\
        \bottomrule
    \end{tabular}
    \caption{Fr{\'e}chet distance (FD) and logit score (LS) results on US8K and TUT data sets. DAG models were generated with sampling parameters $\alpha=2$, $N=100$, $\gamma=2$ for US8K, and $\gamma=0$ for TUT. Metrics computed with OpenL3 at 48\,kHz.}
    \label{tab:mainresults}
\end{table*}

\begin{figure*}[h!]
	%\centering
	\hfill
	\includegraphics[width=0.48\linewidth]{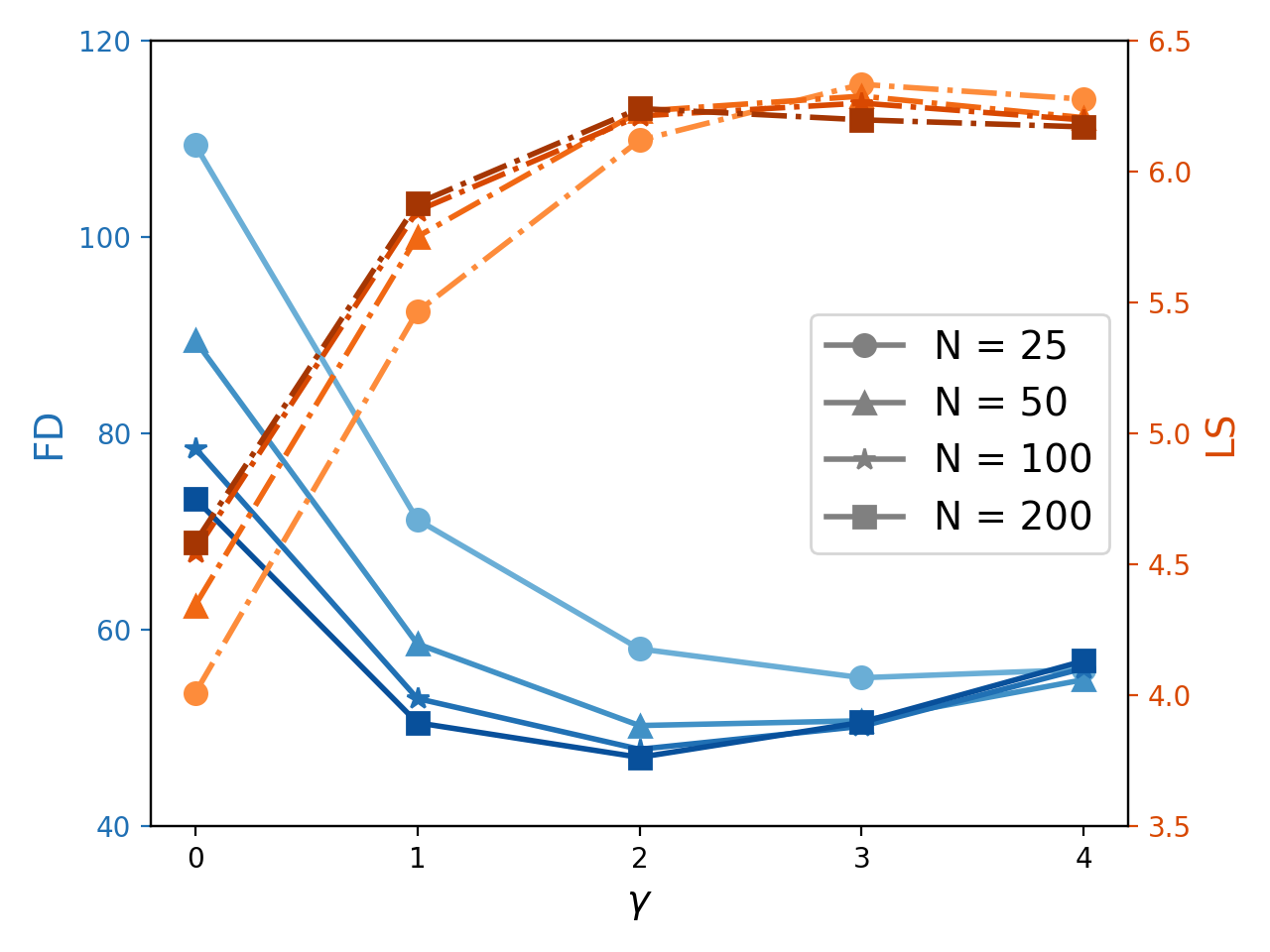}
	\hfill
	\includegraphics[width=0.48\linewidth]{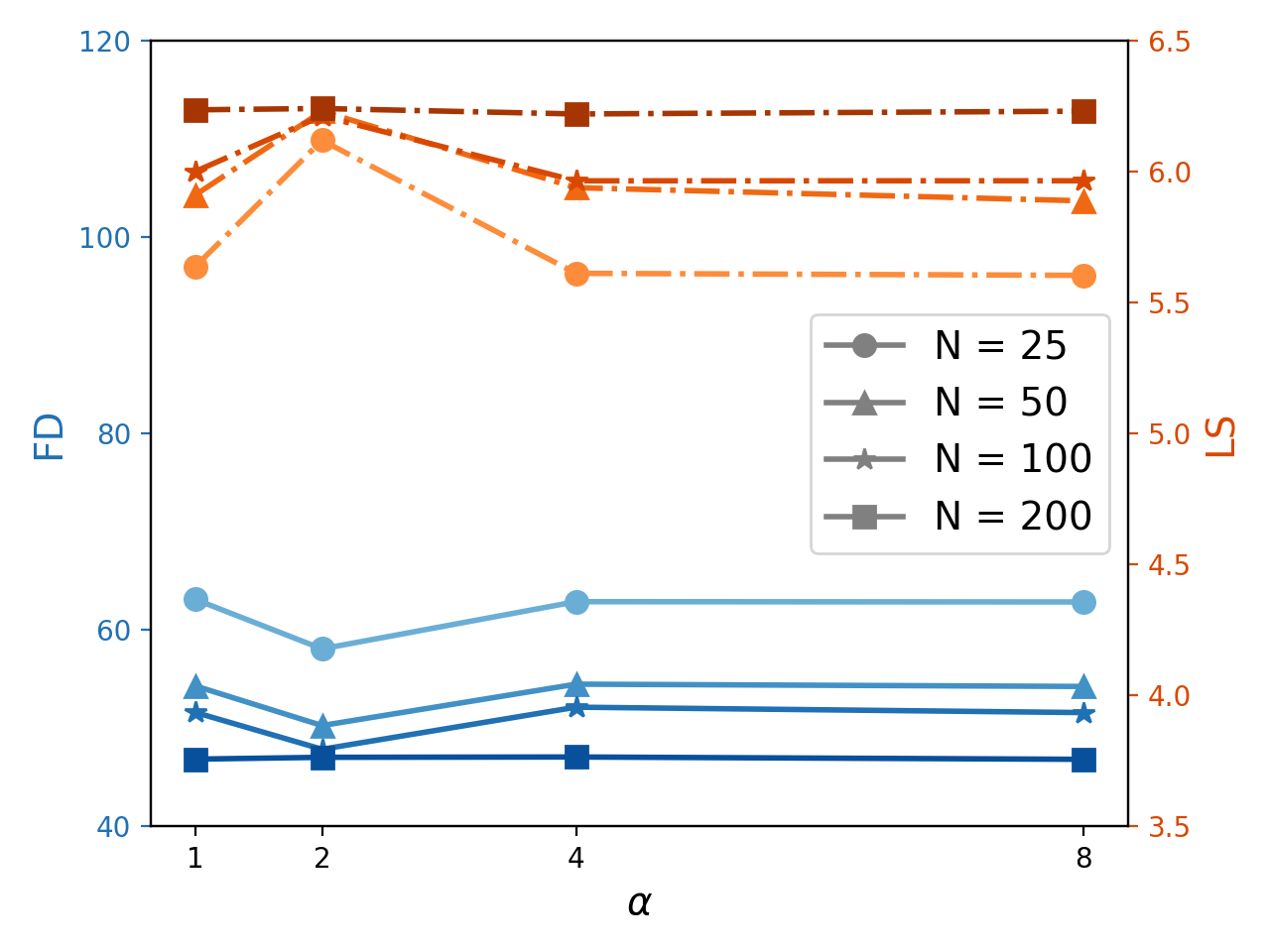} % Placeholder; to be replaced with the alpha plot
	\hfill
	\caption{DAG48 performance as a function of the classifier-free guidance weight $\gamma$ using $\alpha=2$ (left) and the sampling hyper-parameter $\alpha$ using $\gamma=2$ (right) for different number of sampling steps $N$: Fr\'echet distance (FD) and logit score (LS) on US8K dataset. Metrics computed with OpenL3 at 48\,kHz. }
	\label{fig:fd_vs_gamma_us8k}
\end{figure*}

%\begin{table}[h!]
%\centering
	%\resizebox{\columnwidth}{!}{%	
	%\begin{tabular}{l c c c c}
	%	\toprule
	%	\textbf{Model} & \textbf{FD}~~$\downarrow$ & \textbf{LS}~~$\uparrow$ & \textbf{SR [kHz]} & \textbf{params. [M]}\\ 
	%	\midrule
	%	Real22k & $72.60$ & $5.31$ & 22.05 & - \\
	%	SampleRNN & $237.04$ & $2.32$ & 22.05 & 42 \\
	%	PixelSNAIL & $115.12$ & $3.08$ & 22.05 & 107 \\
	%	\midrule
	%	DAG22 & $94.67$ & $5.16$ & 22.05 & 22 \\
	%	DAG22TR & $85.46$ & $5.11$ & 22.05 & 22 \\
	%	DAG48 & $53.03$ & $5.92$ & 48.00 & 22 \\
	%	\bottomrule	
	%\end{tabular}
    %}
	%\caption{Fr{\'e}chet distance (FD) and logit score (LS) results on US8K. \joan{Add ``evaluation done at 48\,kHz''} \joan{I don't think the second decimal matters (to remove)} \joan{Maybe we can add a column with sampling rate and another one with number of parameters} \joan{I think I'd remove Real16k. What does it bring?} }
	%\label{table:us8k}
%\end{table}

%\begin{table}[h!]
%\centering
	%\resizebox{\columnwidth}{!}{%	
	%\begin{tabular}{l c c}
	%	\toprule
	%	\textbf{Model} & \textbf{FD}~~$\downarrow$ & \textbf{LS}~~$\uparrow$ \\ 
	%	\midrule
	%	Real16k & $80.78$ & $0$ \\
	%	SampleRNN16k & $0$ & $0$ \\
	%	\midrule
	%	Real22k & $78.96$ & $0$ \\
	%	SampleRNN & $139.38$ & $1.00$ \\
	%	PixelSNAIL & $144.27$ & $0$ \\
	%	PixelSNAIL-US8KH & $107.31$ & $4.15$ \\
	%	DAG22 G0 & $78.77$ & $4.43$ \\
	%	\midrule
	%	DAG48 G0 & $37.71$ & $6.34$ \\
	%	\bottomrule	
	%\end{tabular}
    %}
	%\caption{Fr{\'e}chet distance and logit score results on TUT. \joan{Same comments as in Table 1} }
	%\label{table:us8k}
%\end{table}

\subsection{Effect of Guidance and Sampling Hyper-parameters}

Figure~\ref{fig:fd_vs_gamma_us8k} shows DAG48 performance when changing the classifier-free guidance weight $\gamma$ (left), as well as the sampling hyper-parameter $\alpha$ (right). Regarding $\gamma$, its increase translates into a quick improvement of the synthesis quality, especially with a low amount of steps in the Langevin sampling $N$. This means that a model with $50$ steps can already get a competitive performance with $\gamma = 2$, which reduces substantially the computational cost when compared to $N=200$. Note that when $N$ is increased, a higher guidance weight degrades the quality, and that this degradation is faster with longer sampling processes. The fact that the LS curves saturate with $\gamma = 4$ but FD increases may indicate a loss in diversity at the expense of more realism, something intrinsic to the classifier-free guidance technique~\cite{ho2022guidance}. 
Regarding $\alpha$, we observe a trend where the choice of a particular value seems to be less critical, especially if we can afford to sample with $N=200$ steps. Nevertheless, a factor $\alpha=2$ works well overall, and is of importance when we want to reduce the sampling computational complexity with $N<200$.

\subsection{Additional Samples and Qualitative Analysis}
\label{subsec:add_results}

It is observed that the TUT and US8K datasets contain mostly noisy ambiance with some target sound events happening in the background, which makes it difficult to identify if some low-quality synthesis samples are due to the training data or a model limitation. Therefore, we use additional clean foreground sounds as training data to further experiment with DAG's modeling capability and generation quality. This data is an in-house collection of carefully-picked high-quality samples from Freesound\footnote{\url{https://freesound.org}}, including wind, rain, river, waves, fire, footsteps, applause, horses, and piano music. This selection aims at covering a wider range of signal characteristics, from modulated noise and sinusoids to transients. 

The resulting synthesis samples are promising: the attacks of footsteps, fire crackling or piano sound crispy; the noisy wind, rain, or waves sound naturally smooth; the individual bubbling in river and the individual clapping in crowd applause can be heard clearly. Some generated examples are available for listening on the project website\footnote{\url{https://diffusionaudiosynthesis.github.io}} as a demonstration of DAG's great potential to synthesize general audio with quality and diversity. Moreover, we show some results of a simple audio style transfer approach enabled by DAG's modeling strategy. Here, we inject a normalized input signal $\ve{y}$ before the first sampling step, such that $\ve{x}_1=\ve{y} +\sigma_1\ve{z}_1$ (see equation~\ref{eq:sampling}).
%We can inject an input signal $\ve{y}$ scaled by a factor $\phi\in (0, 1]$ before the first sampling step, such that $\ve{x}_1=\phi\ve{y} +\sigma_1\ve{z}_1$ (see equation~\ref{eq:sampling}).

\section{Conclusions}
\label{sec:conclusions}

In this work, we propose the diffusion audio generator, a full-band general audio synthesis model based on score-based diffusion. With this, we tackle the source-agnostic audio synthesis problem with an end-to-end solution capable  of outperforming the state of the art in a label-to-audio setup for two datasets that mix environmental and impulsive sounds. Moreover, we investigate the effectiveness of recent sampling techniques for diffusion models, like classifier-free guidance, showing that it can greatly benefit the performance of our model whilst reducing the amount of sampling steps needed to achieve a certain level of quality.

\section{Acknowledgements}
\label{sec:ack}

We would like to thank Giulio Cengarle for fruitful guidance on signal processing methods.

%\end{minipage}
%
%\begin{minipage}[b]{.48\linewidth}
%  \centering
%  \centerline{\includegraphics[width=4.0cm]{image3}}
%  \vspace{1.5cm}
%  \centerline{(b) Results 3}\medskip
%\end{minipage}
%\hfill
%\begin{minipage}[b]{0.48\linewidth}
 % \centering
%  \centerline{\includegraphics[width=4.0cm]{image4}}
%  \vspace{1.5cm}
%  \centerline{(c) Result 4}\medskip
%\end{minipage}
%
%\caption{Example of placing a figure with experimental results.}
%\label{fig:res}
%
%\end{figure}

%\vfill
%\pagebreak

\vfill\pagebreak

%\section{REFERENCES}
%\label{sec:refs}

\bibliographystyle{IEEEbib}
\bibliography{strings,refs}

\end{document}